\documentclass[useAMS,usenatbib]{mn2e}

\usepackage{graphicx}
\usepackage{natbib}
\usepackage{subfloat}
\usepackage{amssymb}
\usepackage{color}

\title[The Origin of Chaos in the Orbit of Comet 1P/Halley]{The Origin of Chaos in the Orbit of Comet 1P/Halley}
\author[T.~C.~N.~Boekholt, F.~I.~Pelupessy, D.~C.~Heggie and S.~F.~Portegies~Zwart]{T.~C.~N.~Boekholt$^{1}$\thanks{E-mail: boekholt@strw.leidenuniv.nl (TB); pelupes@strw.leidenuniv.nl (IP); d.c.heggie@ed.ac.uk (DH); spz@strw.leidenuniv.nl (SPZ)}, F.~I.~Pelupessy$^{1}$\footnotemark[1],   D.~C.~Heggie$^{2}$\footnotemark[1] and S.~F.~Portegies~Zwart$^{1}$\footnotemark[1]\\ 
$^{1}$Leiden Observatory, Leiden University, PO Box 9513, 2300 RA, Leiden, The Netherlands \\
$^{2}$School of Mathematics and Maxwell Institute for Mathematical Sciences, University of Edinburgh, King's Building, Edinburgh EH9 3{FD}, UK}
 
\def\gtorder{\mathrel{\raise.3ex\hbox{$>$}\mkern-14mu
             \lower0.6ex\hbox{$\sim$}}}
\def\ltorder{\mathrel{\raise.3ex\hbox{$<$}\mkern-14mu
             \lower0.6ex\hbox{$\sim$}}}
\def\msun{M_\odot}
\def\omega{f}

\begin{document}

\date{Accepted Year Month Day. Received Year Month Day; in original form Year Month Day}

\pagerange{\pageref{firstpage}--\pageref{lastpage}} \pubyear{2015}

\maketitle

\label{firstpage}


\begin{abstract}

According to \citet{2015MNRAS.447.3775M} the orbit of comet 1P/Halley is chaotic with a surprisingly small Lyapunov time scale of order its orbital period. In this work we analyse the origin of chaos in Halley's orbit and the growth of perturbations, in order to get a better understanding of this unusually short time scale. 
We perform N-body simulations to model Halley's orbit in the Solar System and measure the separation between neighbouring trajectories. To be able to interpret the numerical results, we use a semi-analytical map to demonstrate different growth modes, i.e. linear, oscillatory or exponential, and transitions between these modes.  
We find the Lyapunov time scale of Halley's orbit to be of order 300 years, which is significantly longer than previous estimates in the literature. This discrepancy could be due to the different methods used to measure the Lyapunov time scale. A surprising result is that next to Jupiter, also encounters with Venus contribute to the exponential growth in the next 3000 years.   
Finally, we {note} 
an interesting application of the sub-linear, oscillatory growth mode to an ensemble of bodies moving through the Solar System. { Whereas in the absence of encounters with a third body the ensemble spreads out linearly in time, the accumulation of weak encounters can increase the lifetime of such systems due to the oscillatory behaviour.}  

\end{abstract}




\section{Introduction}\label{Sec:Introduction}

Comet 1P/Halley (hereafter just Halley) has regained considerable interest recently, because of the importance to understand the stability of trajectories in the Solar System, and in planetary systems in general. Small variations in Halley's time of sighting over the last millennium have alerted astronomers to the possible chaotic nature of the comet's orbit \citep{1988PAZh...14..357V}. Its chaoticity has indeed been verified in several studies, by both analytical \citep[e.g.][]{1989A&A...221..146C, 2007IAUS..236...15S, 2014arXiv1410.3727R} and numerical methods \citep{2015MNRAS.447.3775M}.  
The Lyapunov time scale for Halley's orbit has been determined to be on the order of its orbital period  or less, i.e. {$\ltorder$76} years \citep{2015MNRAS.447.3775M}, with a {lower bound} 
of 34 years \citep{2007IAUS..236...15S}.  Th{e} 
unusually short time scale {of \citet{2015MNRAS.447.3775M}} stimulated our curiosity about the chaotic nature of Halley's orbit, specifically the origin of its chaos and its short Lyapunov time scale. 

In comparison, the Lyapunov time scale of the Solar System, i.e. the Sun and the planets, is about 5 Myr \citep{1989Natur.338..237L}, which is much longer than the orbital periods of the planets. The origin of this chaos is therefore sought in secular resonances, {whose periods} 
are typically on the order of many orbital periods \citep{1990Icar...88..266L}. On the other hand, a recent study of the exoplanetary system Kepler-36 revealed a Lyapunov time scale of merely 10 years \citep{2012ApJ...755L..21D}. This is an extremely short time scale in an absolute sense, but still a few hundred times the orbital periods of the two planets in the system, which are 13.8 and 16.2 days \citep{2012Sci...337..556C}. The origin of chaos in this Kepler system is found to be in the interaction between two {mean-motion} 
resonances, specifically the 6:7 and the 29:34 resonances \citep{2012ApJ...755L..21D}. 

{ One possible explanation for the origin of Halley's short Lyapunov time scale may be the overlap in orbital resonances corresponding to integer p:1 ratios between Halley's orbit and the Sun-Jupiter system \citep{2015ApJ...799....8S}. An alternative explanation considers strong deflections of Halley during each of its orbital periods.} This is similar to the chaoticity of the Jupiter family of short period comets, where close encounters with the planets cause a short lived, but significant growth of perturbations, whereas in between encounters the growth is linear \citep{1998CeMDA..70..181T}. On the other hand the Lyapunov time {scale of these objects} is of order 10 orbital periods, which is still inconsistent with the result reported for Halley.

The aim of this study is to understand the origin of chaos in the orbit of Comet Halley and its associated Lyapunov time scale. 
To do this we revisit the problem of Halley's encounters with the planets
using precise N-body calculations and quantitative analyses, taking
special care to analyse the contributions from each of the planets in the Solar System.
In Sec.~\ref{Sec:Encounter} we analyse the growth of an initial perturbation in Halley's orbit due to an encounter with a planet, in order to determine which planets contribute {most
} to Halley's chaoticity. 
In Sec.~\ref{Sec:Exp1} we study the effect of multiple encounters using a map similar to those in  \citet{1989A&A...221..146C} and \citet{ 2014arXiv1410.3727R}, which uses kick-functions to model the perturbations on Halley's orbit. The aim of this semi-analytical model is to understand the underlying mechanism for exponential growth and transitions in the rate of divergence. In Sec.~\ref{Sec:Exp2} we use precise N-body integrations of Halley's orbit to accurately measure the rate of divergence between neighbouring solutions, and to measure the Lyapunov time scale for exponential growth.  {Sections \ref{Sec:Discussion} and \ref{Sec:Conclusions} include further discussion and summarise our main conclusions, respectively.}


\section{Encounters between Halley and a Planet}\label{Sec:Encounter}

If we regard the two-body system consisting of the Sun and Halley, and introduce a small perturbation in the orbit of Halley, then this perturbation will grow {approximately} linearly in time.  Due to the slight difference in the orbital period of Halley, the fiducial and perturbed trajectories slowly get out of phase until the perturbation has grown to the size of the orbit. In reality, on top of this steady growth, there is also the effect of close encounters with the planets, which can cause jumps in the {rate of} growth.  

In view of the conjecture that the chaoticity of the orbit of
Halley is determined by close encounters with the planets, we first make some numerical
and order-of-magnitude estimates of the expected effect. It is usually assumed that Jupiter is the most important planet to cause disturbances being the most massive planet, but we will show that other planets contribute as well as the growth of a perturbation is specified by both the mass and the distance to the planet \citep{Newton:1687}. 

{\subsection{Numerical Estimates}\label{sec:numerical}}

\begin{figure}
\centering
\begin{tabular}{l}
\includegraphics[height=0.65\textwidth, width=0.45\textwidth]{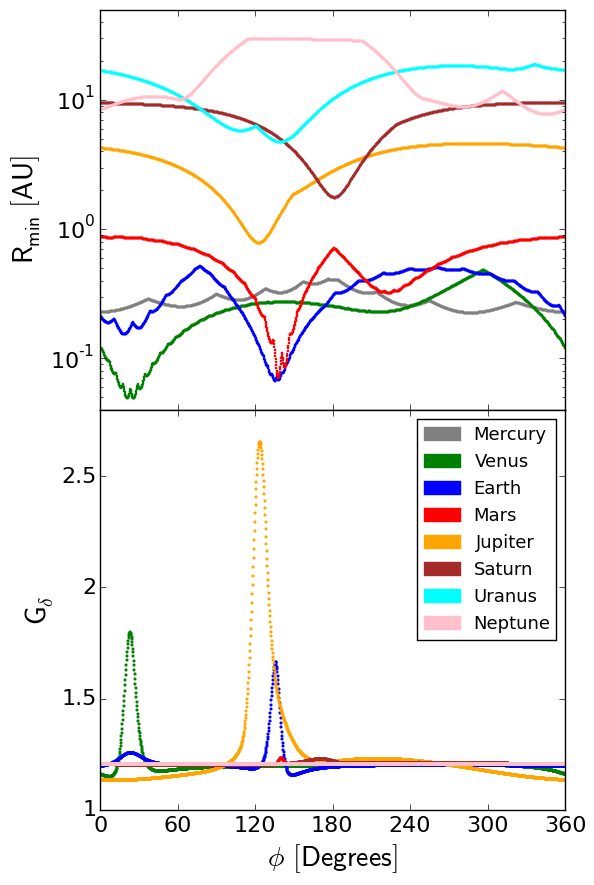} \\
\end{tabular}  
\caption{In the top diagram we plot the closest distance between Comet Halley and a planet during one orbital period of Halley as a function of the initial orbital phase of the planet. In the bottom diagram we plot the growth factor of an initial perturbation in the orbit of Halley after one orbital period, again as a function of initial orbital phase of the planet. We observe that Jupiter (highest peak in bottom panel), but also Venus (high peak between $\phi = 0^{\circ}$ and $60^{\circ}$) and Earth (smaller peak near Jupiter's peak) give rise to the largest magnification factors, and that it correlates to close encounters.}
\label{fig:planet_encounter}
\end{figure}

We perform a series of three-body experiments, consisting of the Sun, Halley and one planet, as we investigate the effect of each planet independently. Whether Halley experiences a close encounter with the planet depends on the initial orbital phase of that planet, $\phi$. Therefore, we systematically vary the value of $\phi$ and measure the growth of {the} 
perturbation in Halley's orbit {over one orbital period of Halley, starting from an initial value $\delta_0$.  Here the perturbation is measured by the Euclidean norm of the vector perturbation in {position}, i.e. the expression
{
  \begin{equation}
    \delta = \sqrt{\delta x^2 + \delta y^2 + \delta z^2},\label{eq:norm}
  \end{equation}
}
\noindent where the unit of {length} 
is 1\,AU. 
The growth of the perturbation is measured by the factor} $G_{\delta} = \delta_f/\delta_0$, 
{where} $\delta_f$ 
{is the} perturbation {after one period.} 
A difference in the growth factor is then purely caused by a difference in the encounter.    

Currently, Halley is close to its aphelion, which is an appropriate initial state {for} our experiments. We take the initial conditions from the JPL Horizons database \footnote{http://ssd.jpl.nasa.gov/, JDCT $=$ 2456934.5 $=$ A.D. 2014-Oct-04 00:00:00.0000 (CT)}, {in the {barycentric} frame. For each}  
planet, we {systematically} generate 1440 realisations of the planet's initial true anomaly between 0 and 360 degrees in steps of 0.25 degree{, while keeping Halley fixed}.  We generate {both this} 
fiducial initial condition, and a perturbed initial condition, where we introduce an offset of $10^{-6}$ AU in the {$x$}-coordinate of Halley. To make sure that numerical artefacts are negligible, we use the arbitrary-precision N-body code \texttt{Brutus} \citep{2015ComAC...2....2B} with a Bulirsch-Stoer tolerance of $10^{-10}$ and a word-length of 72 bits (or about 22 decimal places). We integrate the systems for one orbital period of Halley, i.e. 76 years, and evaluate the magnitude of the perturbation in {position} 
in order to determine the growth factor $G_{\delta}$. 

We show the results of this experiment in Fig. \ref{fig:planet_encounter}. In the top panel we plot the closest approach between Halley and the planet during one orbital period of Halley, $R_{min}$, as a function of the initial orbital phase of the planet, $\phi$. We generally observe that there are small intervals in $\phi$ where close approaches occur. For example, Jupiter (orange curve, starting at $R_{min}=4$\,AU at $\phi=0^{\circ}$), has its smallest value {$R_{min}=0.78$\,AU} at $\phi=123^{\circ}$. Venus on the other hand (green, bottom curve at $\phi=0^{\circ}$), has its smallest value {$R_{min}=0.049$\,AU} at $\phi=16^{\circ}${, while for Earth $R_{min}=0.067$\,AU}.     
{The small scale oscillations present in the curves are a numerical artefact, caused by the constant time intervals at which we evaluate $R$, so that sometimes the true $R_{min}$ is slightly missed. }

In the bottom diagram of Fig. \ref{fig:planet_encounter} we measure the growth factor of an initial perturbation, $G_{\delta}$, as a function of the initial orbital phase of the planet, $\phi$. 
We observe that Jupiter (orange, highest-peaked curve), Venus (green, peak between $\phi=0^{\circ}$ and $60^{\circ}$) and Earth (blue, smaller peak near Jupiter's peak), are the dominant planets, with magnification factors ranging from {1.66} for Earth, to {1.80} for Venus and {2.66} for Jupiter. The relative importance of Venus and Earth can be understood by noting that the inclined orbit of Halley crosses the orbital plane of the planets {close} 
 to the orbit of Venus. 

If we compare the diagrams in Fig. \ref{fig:planet_encounter}, we observe that the highest {significant} magnification factor{s} 
correlate 
with the closest approach between Halley and the {corresponding} planet. Therefore, the effect of the planets on the growth of a perturbation in Halley's orbit is to give {a} short lived but significant kick to the perturbation growth, but {mainly} 
in the cases of closest approach. It seems possible then, that if Halley experiences a series of rather weak encounters with Jupiter, its chaoticity can be fuelled by Venus or Earth instead. 

Using the maxima of the growth factors in Fig.~\ref{fig:planet_encounter} (lower panel), we can obtain a rough estimate for the lower limit of the Lyapunov time scale of Halley's orbit. We assume a constant sequence of encounters between Halley and the planets at equal time intervals of an orbital period, $P$. Also, each encounter has the maximum strength, i.e. maximum growth factor $G_{\delta}$. This results in the following expression: $\delta(P) = \delta(0) e^{\lambda P}$, {where $\lambda$ is the (maximum) Lyapunov exponent. Thus} $G_{\delta} \equiv \delta(P)/\delta(0) = e^{\lambda P}$, $\lambda = \log G_{\delta}/P$, or in terms of the Lyapunov time{,} $\tau = 1/\lambda = P/\log G_{\delta}$. Filling in the maximum magnification factors mentioned above, we obtain {149}\,yr for Earth, {128}\,yr for Venus and 77\,yr for Jupiter. { In the case that Halley's chaoticity were to be dominated by close encounters,} these results { suggest} that the actual Lyapunov time scale must be {substantially} larger than the orbital period of Halley{, as these closest encounters are uncommon.}

\subsection{Order-of-magnitude Estimates}\label{sec:analytical-estimates}

{
Surprising though it may be that planets with a mass of order $10^{-3}$ times that of Jupiter nevertheless produce comparable effects on the motion of Halley, simple estimates demonstrate the compensating effect of the distance of closest approach, as we now show.}

{
We first consider the effect of an encounter on the fiducial orbit.  If the distance of closest approach is $d$, then the change in velocity of Halley may be estimated as
\begin{equation}
\Delta v_H \simeq \frac{2Gm_p}{dv_r},
\end{equation}
where $m_p$ is the mass of the planet and $v_r$ is the relative speed of Halley and the planet at the time of the encounter.  (The factor 2 comes from treating the encounter impulsively \citep[][eq.(1.30)]{2008gady.book.....B}.)  Multiplying by $v_H$, we estimate the change in (specific) energy of Halley, and hence estimate {the relative change in $a$, the semi-major axis of Halley.  Assuming that the relative}
change in apocentric distance $Q$ is comparable, we find
\begin{equation}
\Delta Q \simeq 4Q\frac{m_p}{\msun}\frac{v_H}{v_r}\frac{a}{d},\label{eq:deltaQ}
\end{equation}
where $\msun$ is the solar mass. 
Note that the factor $v_H/v_r$ will be of order 1/2, as the orbit of Halley is retrograde, and Halley and the planet have comparable speeds.}

{
Now we consider the perturbed orbit, which starts at apocentre with a small perturbation $\delta_0$ in position.  When the comet reaches the vicinity of a planet with orbital radius $r_p$, this initial perturbation will have increased by the ratio of the speed of Halley at apocentre and at the planetary radius.  Therefore the perturbation in position, which will also be approximately the perturbation in the distance of closest approach, is $\delta d\simeq \delta_0\sqrt{Q/r_p}$, where we have estimated speeds by using a parabolic approximation for the motion of Halley.  It follows from Eq.~(\ref{eq:deltaQ}) that the perturbation in position at the next apocentre is 
$\delta_f \simeq \delta Q \simeq \delta d.\partial\Delta Q/\partial d$, i.e.
\begin{equation}
\delta_f \simeq 4\delta_0\frac{Q}{a}\frac{m_p}{\msun}\frac{v_H}{v_r}\left(\frac{Q}{r_p}\right)^{1/2}\frac{a^2}{d^2}.
\end{equation}
Estimating $v_H/v_r\simeq1/2$, as noted above, and minimum values of $d$ noted in Sec.~\ref{sec:numerical}, we readily estimate { $G_{\delta} \simeq 8.9$, $5.0$ and $5.1$ for Venus, Earth and Jupiter respectively, while the values for Mars and Saturn are an order of magnitude smaller, $0.4$ and $0.2$ respectively.} While these { simple} estimates { somewhat overestimate} 
the values measured numerically in Sec.~\ref{sec:numerical}{, especially for Venus,} 
they { do} explain why these {three} planets give comparable (maximum) contributions to the growth of perturbations in Halley's orbit {and dominate compared to those from other planets}.  }

\section{The Onset of Exponential Divergence}\label{Sec:Exp1}

{ In this section we investigate the effect of multiple encounters, i.e. an encounter history, on the growth of a perturbation in Halley's orbit. Using a semi-analytical map we will demonstrate that there are three growth modes, i.e. exponential, oscillatory and linear, and that transitions between them correlate with close encounters. This analysis is crucial for interpreting the numerical results in the next section.}

\subsection{Map for Changes in Orbital Frequency}\label{mapeq}   


{ We construct a map similar to \citet{1989A&A...221..146C} of the time evolution for the orbital frequency of Halley and the phase of the planet at closest distance (here we regard Jupiter).} 
{The map is given by
\begin{equation}\label{MAPEq:2}
\phi_{n+1} = \phi_n + 2 \pi \left(\frac{\omega_J}{\omega_n}\right),
\end{equation}
\begin{equation}\label{MAPEq:omega}
  \omega_{n+1} = \omega_n + K(\phi_{n+1}),
\end{equation}
where $\phi_n$ is the phase (i.e. longitude) of Jupiter at the $n$th perihelion passage, $\omega_n$ is the frequency of Halley after the $n$th perihelion passage, $\omega_J$ is the (constant) frequency of Jupiter { and $K$ is the kick function that models the effect of the encounter.}  The times can be obtained recursively from 
\begin{equation}\label{MAPEq:1}
t_{n+1} = t_n + \frac{1}{\omega_n}.
\end{equation}
Time is measured in years, $\omega$ in yr$^{-1}$ and semi-major axis, when we need it, in AU.   
 
{ The} tangent map, i.e. the linearisation of the above map, { is} given by 

\begin{equation}\label{MAPEq:3}
\delta \phi_{n+1} = \delta \phi_n - 2 \pi \frac{\omega_J}{\omega_n^2} \delta \omega_n, 
\end{equation}
\begin{equation}\label{MAPEq:4}
\delta \omega_{n+1} = \delta \omega_n + \delta\phi_{n+1}K'(\phi_{n+1}).
\end{equation}
When the right side of Eq.~(\ref{MAPEq:4}) is expressed in terms of $\delta\phi_n$ and $\delta\omega_n$, it takes the form
\begin{equation}
  \delta \omega_{n+1} = \delta \omega_n + \left(\delta\phi_{n} - 2\pi\frac{\omega_J}{\omega_n^2}\delta\omega_n\right)K'(\phi_{n+1}).
\end{equation}
Combining with Eq.~(\ref{MAPEq:3}), we see that 
the matrix of the linearised map is given by 
\begin{equation}
  A = \left(
    \begin{array}{cc}
1 & -2\pi\displaystyle{\frac{\omega_J}{\omega_n^2}}\\
K'(\phi_{n+1}) & 1 - 2\pi\displaystyle{\frac{\omega_J}{\omega_n^2}}K'(\phi_{n+1})
    \end{array}\right).
\end{equation}
This matrix has determinant one, showing that our map is {\sl symplectic} (i.e. area-preserving).  Thus although the variables $\omega,\phi$ are not canonical in the usual sense (energy and phase would be better), the map preserves the main geometrical constraint of a canonical mapping.  


{ The eigenvalues of the matrix are}
\begin{equation}\label{eq:evals}
  \lambda = 1 -\pi\frac{\omega_J}{\omega_n^2}K' \pm \sqrt{\pi\frac{\omega_J}{\omega_n^2}K'\left(\pi\frac{\omega_J}{\omega_n^2}K' - 2\right)},
\end{equation}
where $K' = K'(\phi_{n+1})$.  {Thus}
\begin{equation}
  \lambda \simeq 1 \pm i\sqrt{ \frac{2\pi\omega_JK'}{\omega_n^2}}\label{eq:lkf-evals}
\end{equation}
when $\vert K'\vert \ll 1$.  {These results show} 
that the evolution { of the perturbation growth} is expected to be {one of exponential growth {\sl unless}}
\begin{equation}
  0 < K' < \frac{2f_n^2}{\pi f_J}.
\end{equation}
When $K'$ lies within this range the evolution is expected to be oscillatory and periodic, with a 
period (in years) 
given approximately by
\begin{equation}
  P = \sqrt{\frac{2\pi}{\omega_JK'}}\label{eq:flip-prediction},
\end{equation}
{when $\vert K'\vert \ll 1$.  When $K'=0$ the growth is linear.}

{These remarks about the evolution of the solutions ignore the fact that $K'$ is a function of $\phi_{n+1}$, i.e. the circumstances of a given encounter.  Nevertheless we shall see in the next subsection that quite realistic sequences of encounters result in evolution which can exhibit some aspects of the behaviour we have stated here.}


\subsection{Saw-tooth Kick Function}\label{sec:sawtooth}

\begin{figure*}
\centering
\begin{tabular}{l}
\includegraphics[height=0.45\textwidth, width=0.9\textwidth]{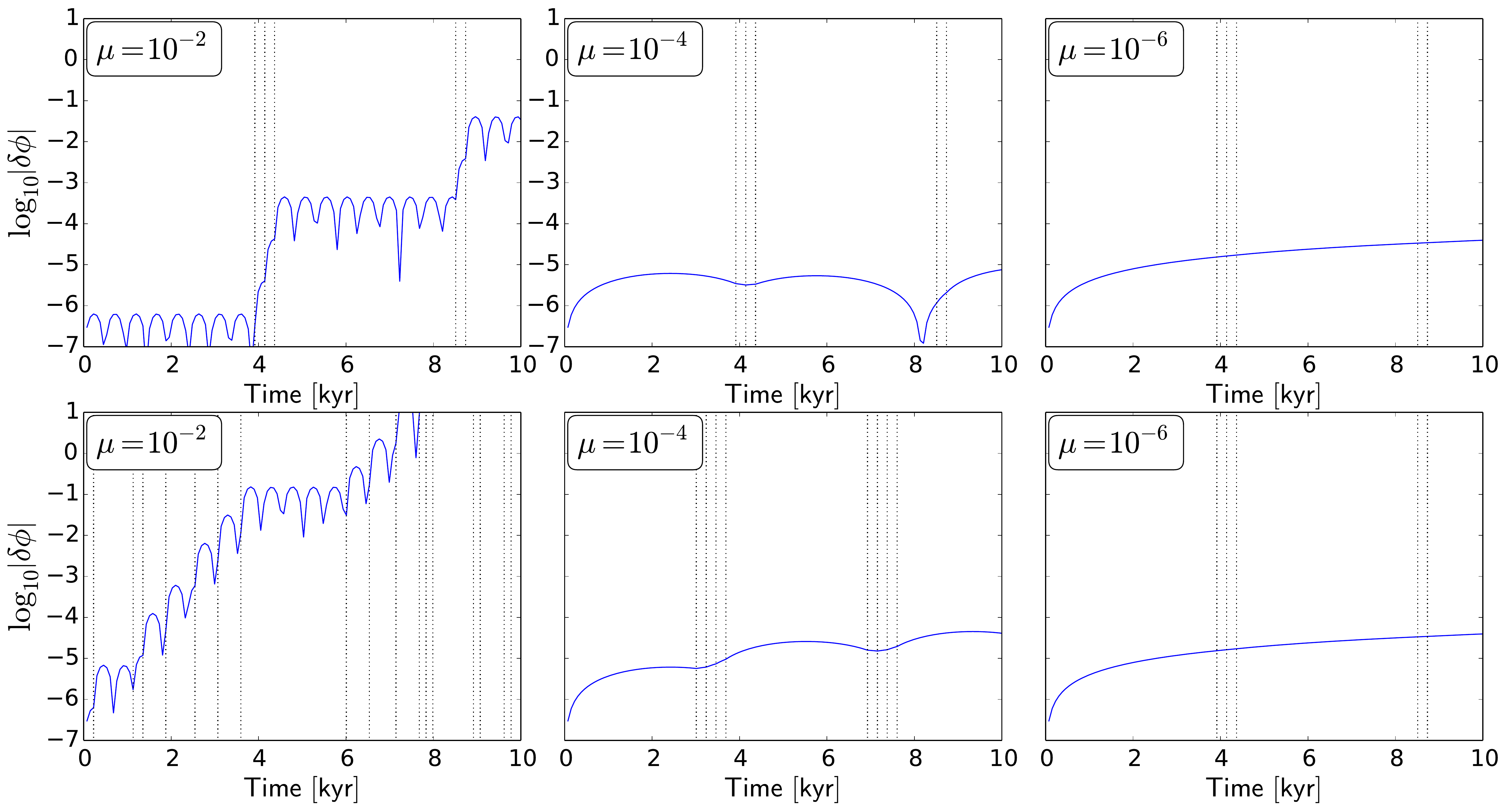} \\
\end{tabular}  
\caption{{The growth of perturbations{, measured by $\vert\delta\phi\vert$,} as a function of encounter strength and history. The encounter strength{,}  quantified by the parameter $\mu$, varies from strong (left column) to weak (right column). The top row is for the map where the orbital period of Halley is kept constant, whereas in the bottom row it is allowed to vary. In each panel the close encounter times are marked by the vertical dotted lines. In the top row, we observe that transitions in the rate of divergence are correlated with sequences of close encounters. In the bottom row, we observe that small perturbations in Halley's orbital period 
{can} increase 
the number of close encounters, which influences the rate of divergence.  {For other initial conditions and parameter values, however, the number may decrease.}}
}
\label{fig:map_sawtooth_overview}
\end{figure*}

{In the previous subsection we considered, in effect, a constant value of $K'$.}
In {the case of} a 
realistic kick function 
both weak and strong encounters {will be} present{, with differing values of $K'$}.  To take this into account we use a{n idealised} saw-tooth kick function{, given by} 

{
\begin{equation}
K(\phi) = -\frac{\mu}{2\pi}(\phi - \phi_w/2), 0 \leq \phi < \phi_w,
\end{equation}
\begin{equation}
K(\phi) = \frac{\mu}{2\pi}\frac{\phi_w}{2\pi-\phi_w}(\phi - \pi + \phi_w/2), \phi_w \leq \phi < 2\pi.
\end{equation}
}




\noindent  { Here $\mu$ represents the strength of the encounters and} $\phi_w$ stands for the {size of the} interval or window in which strong encounters occur, which we take to be 
0.3{,} as estimated from the peaks in Fig. \ref{fig:planet_encounter} (bottom panel)\footnote{{ \citet{1989A&A...221..146C} give, on a different basis, a value which translates to 0.55.}}.   {Note that the relatively small value of $\phi_w$ ensures that the magnitude of $K'$ is much higher inside this window than at other phases.}  {Also, the  values in this window are negative, in agreement with the results of  \citet{1989A&A...221..146C} for Jupiter, when translated to our variables.  It follows from the results of the previous subsection that strong encounters will give rise to exponential growth of perturbations, whereas weak encounters may also give rise to periodic growth, if weak enough.}

{For the sake of illustration, in this subsection we take} the orbital periods of Halley and Jupiter {to be} 
given by $P_h \simeq 75.3$ yr and $P_J \simeq 11.9$ yr respectively.  Note that they are approximately in a 3:19 resonance, {with} $3P_h -19P_j = -0.2$yr, and as a result {Eq.~(\ref{MAPEq:2}) shows that $\phi_{n+3} \simeq \phi_n - 0.11 (\mbox{mod}2\pi)$.  Thus {if the periods remained constant,} a strong encounter would be followed  at intervals of $3P_H$ by two others, and then the pattern would} recur {(possibly with only two strong encounters)} at intervals of about 4.5\,kyr. {In reality however, strong encounters will affect this approximate resonance resulting in either an increase or decrease in the number of strong encounters. To study this effect in more detail, { we present results both for a map where the orbital period of Halley is kept fixed (Fig. \ref{fig:map_sawtooth_overview} top row), and for one where the orbital period varies due to kicks received from the planet (Fig. \ref{fig:map_sawtooth_overview} bottom row).} }

{In the top left panel { of Fig. \ref{fig:map_sawtooth_overview}}, the growth of the perturbation starts out oscillatory { (Eq.~(\ref{eq:flip-prediction}))}. After about 4\,kyr there is a sequence of three close encounters causing a transition onto a faster exponential growth{, as $K' < 0$}. When the sequence of close encounters is over, the oscillatory growth mode is restored. In the other two panels in the top row, the sequences {of close encounters} do not cause {significant} 
exponential growth, because of the smaller value of $\mu$. In the second panel the relatively strong encounters interrupt the oscillatory behaviour {(whose period is several kyr)}, causing {growth} 
again at 4\,kyr.   }

{We observe in each panel{,} in the top row, that the times and number of strong encounters are the same, which is due to the {assumption of} fixed orbital periods. When we allow the orbital period of Halley to vary {(in accordance with Eq.~(\ref{MAPEq:omega}))}, we observe the consistent result that for very weak encounters (bottom, right panel) the encounter sequence is preserved. For larger values of $\mu$ however, the approximate resonance is broken causing a significant increase in the number of strong encounters (17 in the bottom, left panel), { giving rise to a fast exponential growth}. This is more than what would be expected from a purely random sampling of $\phi$, {which for} 
$\phi_w = 0.3$ results in about 6 strong encounters in 10\,kyr {on average}. Th{e} 
assumption {of random sampling} thus seems unjustified and instead there are quasi-resonant sequences which cause a systematic clustering of strong encounters. 
This same mechanism however, can also { (for other choices of initial conditions)} cause a significant {\sl decrease} in the number of strong encounters if the near-resonant sequence keeps missing the strong encounter window, { which would result in a rather slow perturbation growth, i.e. oscillatory or linear.} 

  
\section{N-body Simulations of Halley's Orbit}\label{Sec:Exp2}

\begin{figure}
\centering
\begin{tabular}{c}
\includegraphics[height=0.36\textwidth, width=0.45\textwidth]{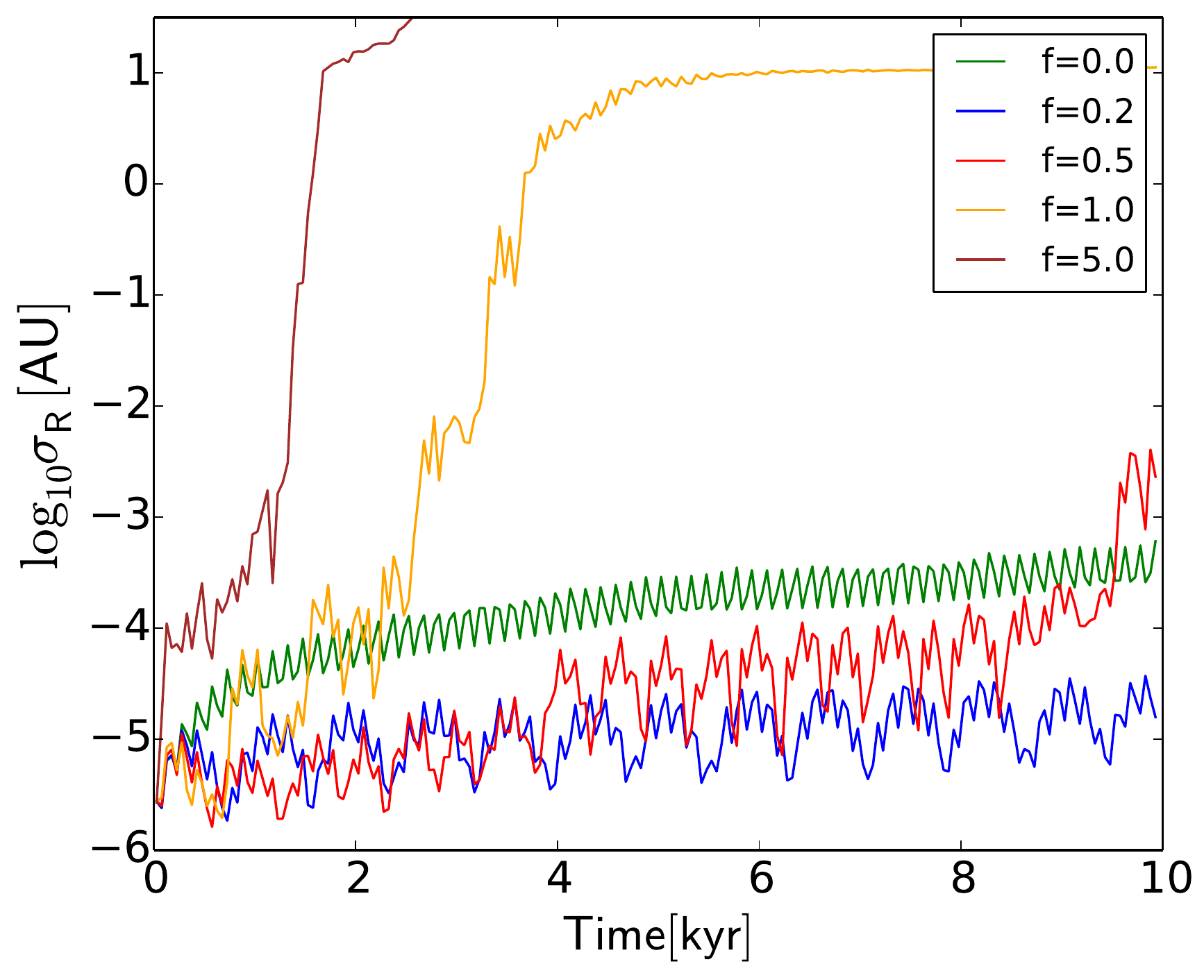} \\
\end{tabular}
\caption{ Growth of the spread in position of an ensemble of Halley-like objects. We vary
the mass of Jupiter by multiplying it by a fraction given in the legend. We reproduce the
linear (green curve), {oscillatory} 
(red and blue curves) and exponential (yellow and brown curves)
growth, depending on the strength of the perturbation. } 
\label{fig:swarm}
\end{figure}

In this section we describe several experiments in which we perform a series of N-body simulations to accurately measure the growth of an initial perturbation in Halley's orbit.  
We model the dynamical evolution of the Solar System according to Newtonian dynamics
, in which the bodies are mathematical point particles. Non-gravitational effects, such as radiation pressure from the Sun, Halley's mass loss due to the interaction with the stellar wind or internal processes, are neglected. This makes our results less realistic, but in order to compare with previous studies, we adopt the same assumptions. The 
non-gravitational perturbations {are} also 
{much} smaller {in magnitude} than the {gravitational forces}  \citep{1998CeMDA..70..181T}, {and we assume that their effect on the generation of chaos is also much smaller}. 
Relativistic effects, especially the orbital precession of Mercury, will also be neglected, since the contribution of Mercury to Halley's chaoticity is very small (see Sec.~\ref{Sec:Encounter}). 

\begin{figure*}
\centering
\begin{tabular}{llll}
\includegraphics[height=0.18\textwidth, width=0.22\textwidth]{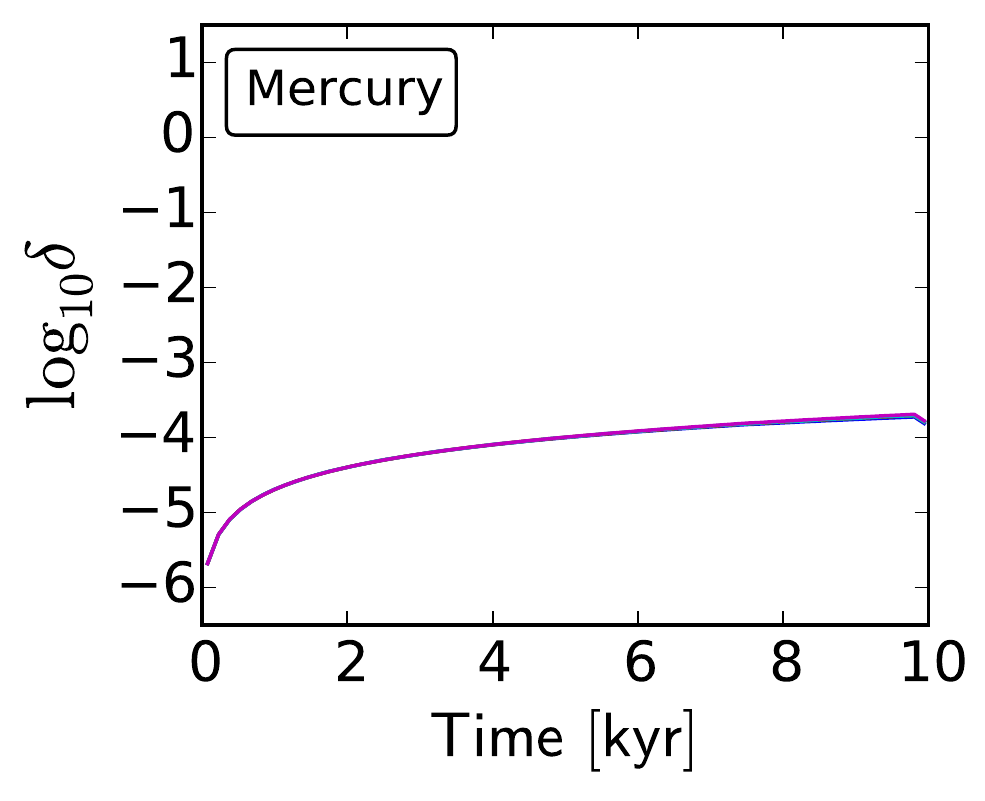} &
\includegraphics[height=0.18\textwidth, width=0.22\textwidth]{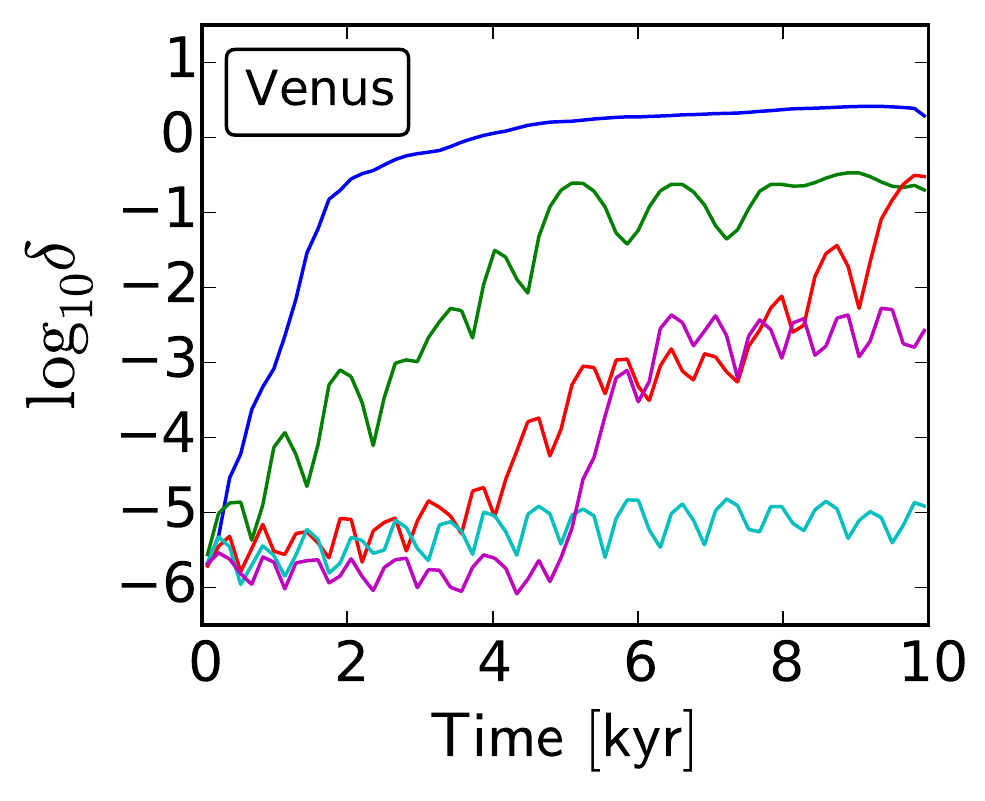} &
\includegraphics[height=0.18\textwidth, width=0.22\textwidth]{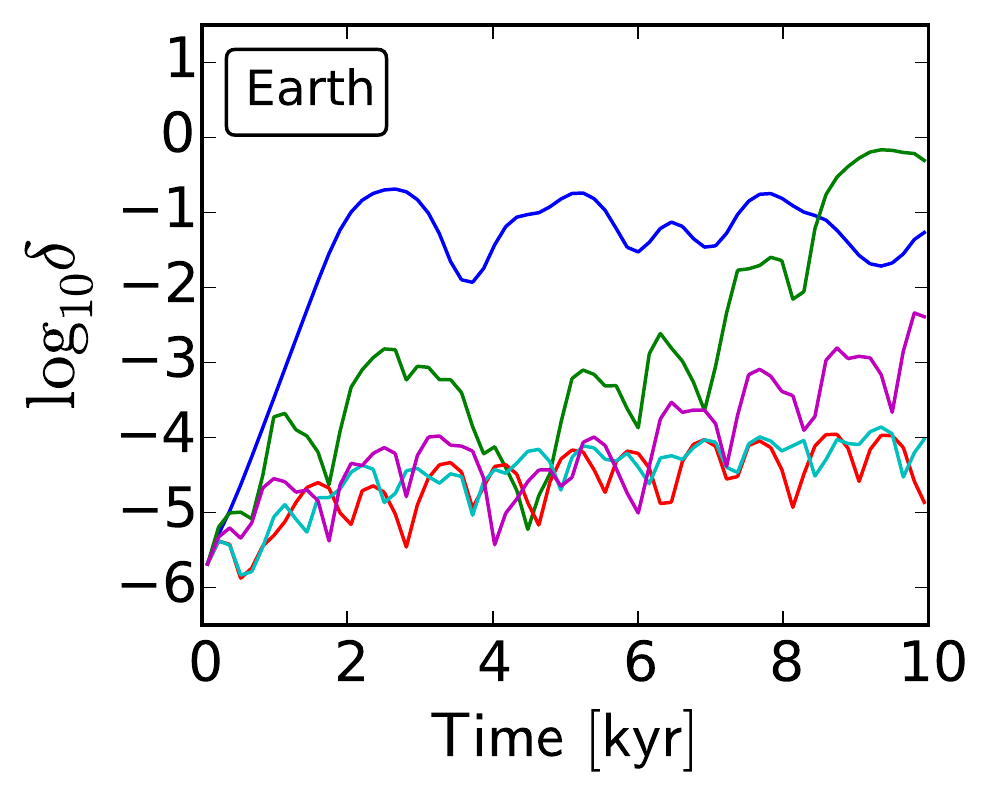} &
\includegraphics[height=0.18\textwidth, width=0.22\textwidth]{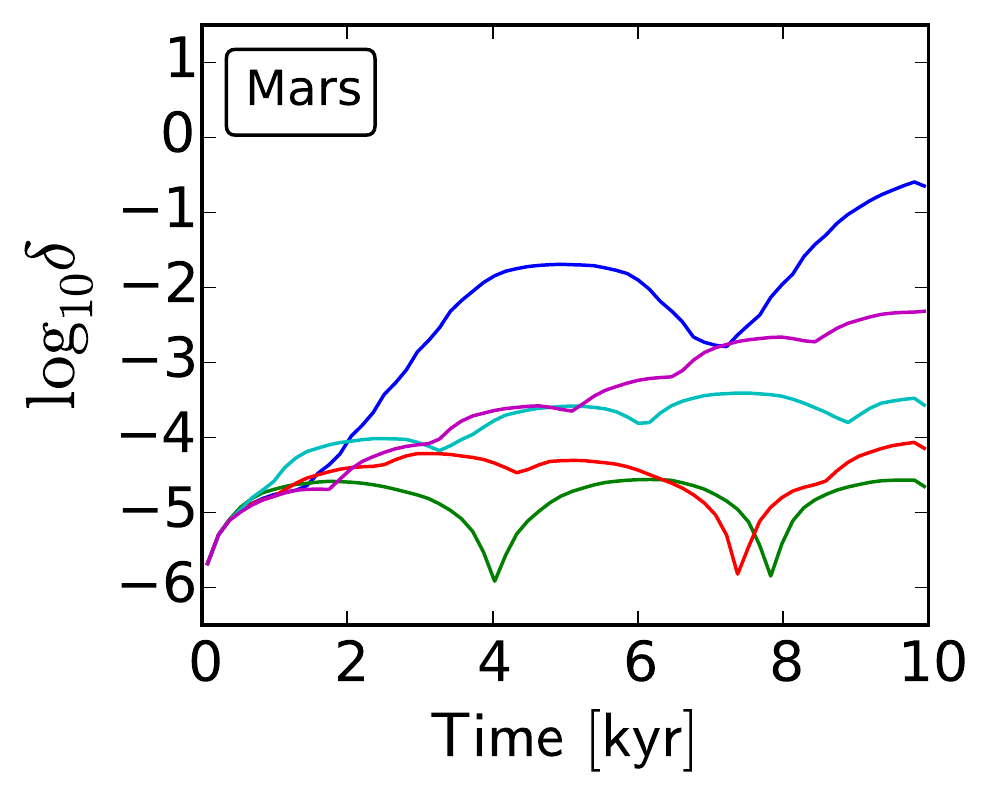} \\
\includegraphics[height=0.18\textwidth, width=0.22\textwidth]{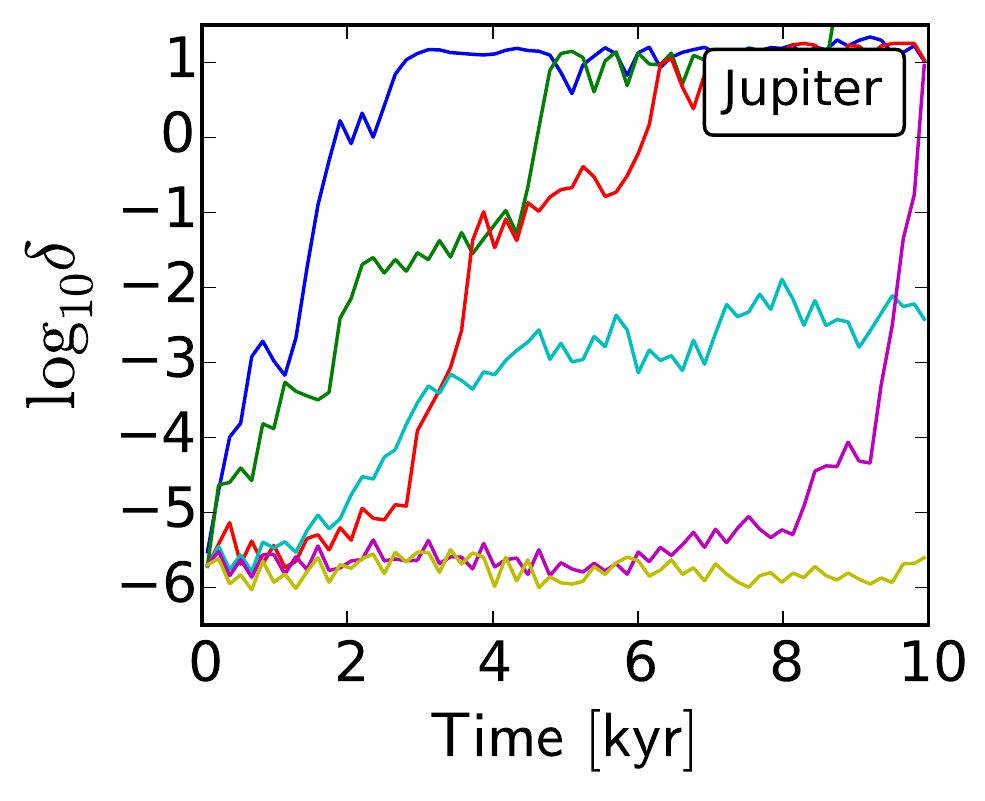} &
\includegraphics[height=0.18\textwidth, width=0.22\textwidth]{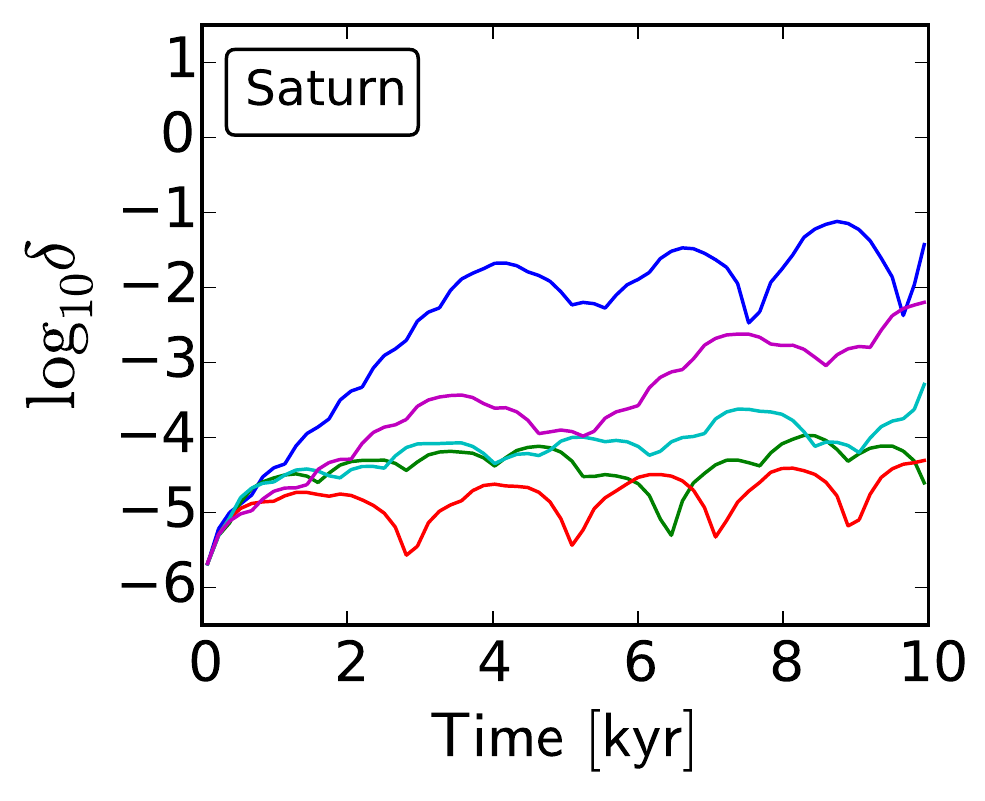} &
\includegraphics[height=0.18\textwidth, width=0.22\textwidth]{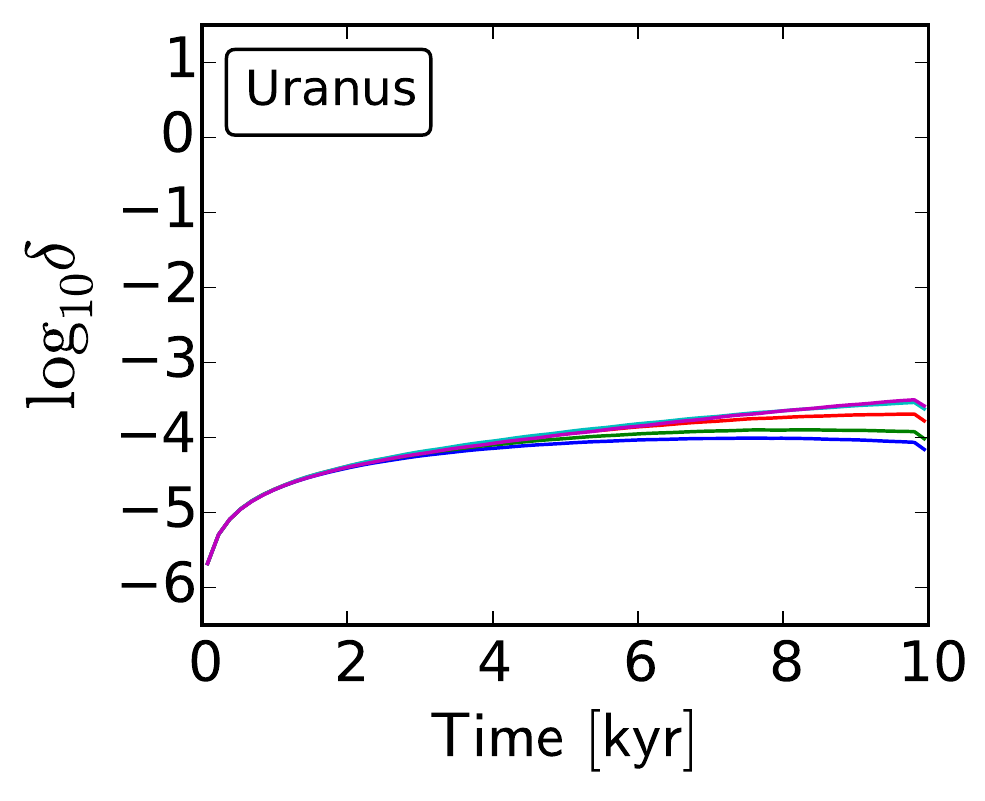} &
\includegraphics[height=0.18\textwidth, width=0.22\textwidth]{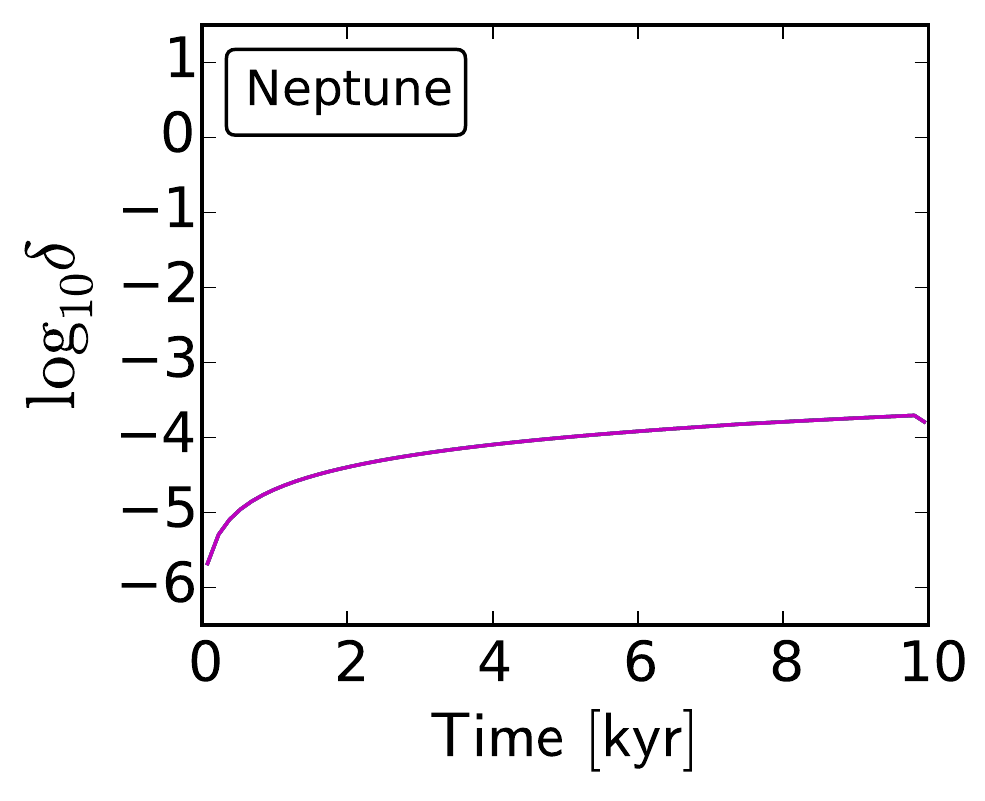} \\
\end{tabular}  
\caption{ Divergence between neighbouring solutions in the $N=3$ Sun, planet and Halley system. We show a subset of solutions to illustrate the different behaviour when we vary the initial orbital phase of the planet around the Sun. As a consequence, in every solution Halley has a different encounter history with the planet. Mercury, Uranus and Neptune do not influence Halley's chaoticity significantly. The other planets are able to cause exponential growth, most notably Jupiter, Venus and Earth.  }
\label{fig:N3_sun_planet_halley_overview}
\end{figure*}

\subsection{Three-body Divergence: Sun, Jupiter and Halley}
\label{Sec:ensemble}


We first {return to the three-body systems of Sec.~\ref{sec:numerical}, studying the effect of}  
a {single} perturbing planet. {
  Those integrations had two limitations, which we aim to remove here.  First, the initial separation between the fiducial and perturbed orbit was of fixed magnitude and direction, and,  second, the integration was followed for only one Halley period.}  
{In order to relax the first limitation,} 
we take an ensemble of a 
hundred Halley-like objects, which are distributed around the fiducial 
initial position in a three-dimensional Gaussian distribution with a 
dispersion of $10^{-6}$ AU. This eliminates any chance effects of preferred 
spatial directions.  {As in Sec.~\ref{sec:numerical}} we only consider the {accelerations} 
due to the Sun and {one planet, for which we take}
Jupiter.  {We also start with the same initial conditions, but with the fiducial initial orbital phase of Jupiter.}   The simulations are done with the \texttt{Huayno} integrator 
\citep{2012NewA...17..711P}.  We choose this integrator instead of \texttt{Brutus} as \texttt{Huayno} is more efficient with larger particle numbers, while still being sufficiently precise.  {As the results of Sec.~\ref{sec:sawtooth} imply that the expected behaviour depends on the strength of the perturbation,}
we {perform different integrations in which we} vary the mass of Jupiter by multiplying it by a factor 
ranging from zero to five.  We measure the spread in the positions of the 
Halley-like objects, i.e. the standard deviation {$\sigma_R$} in the position of the 
ensemble, as a function of time.

We observe in Fig. \ref{fig:swarm} that if the planet has zero mass, we
obtain a linear growth in the dispersion of the positions of the swarm, as
expected from the {model analysis of Sec.~\ref{sec:sawtooth}.  The one new feature is a small-scale oscillation with the period of Halley.}   
For small
Jupiter masses, i.e. a mass smaller than the actual Jupiter mass, we get
a{n oscillatory  behaviour which is, in fact, nearly periodic.  The period is smaller when the perturbation is larger, as expected from Eq.~(\ref{eq:flip-prediction}), but for the larger perturbation (i.e. the case $f=0.5$ in Fig.\ref{fig:swarm}), there appears to be a transition at around 4\,kyr, which is presumably due to one or more particularly close  encounters.}  
Remarkably, when comparing the curve representing $0.2\times M_{jup}$
{to that of $0.5\times M_{jup}$}, we do not observe an increase in the {magnitude of the} growth 
before 4\,kyr, {but} the larger mass {does apparently} 
increas{e} 
the
probability of {Halley} eventually {experiencing} 
a strong interaction.
{A second similar}
strong perturbation 
{also} happens after {about}
$9$\,kyr for {the  case in which the mass of the perturbing planet is} half of Jupiter's mass (red curve).     For heavier {perturber masses}
(i.e. $1\times M_{jup}$ and heavier), we obtain a rather fast
exponential divergence, 
{ but again this behaviour appears to be delayed until the occurrence of 
close encounters after 1 or 2\,kyr.}

Note that the experiment conducted here considers the evolution of an 
ensemble of Halley-like objects. The results would equally apply to a swarm of 
objects (e.g. the result of an asteroid collision or dust emitted from a 
cometary nucleus). The practical effect is that in configurations where the 
orbit 
is affected by 
perturbations {of intermediate strength (illustrated} in the example here 
by {a perturber of mass $0.2\times M_{jup}$}) 
such a swarm {may} 
survive as a coherent group longer than might be expected from the 
linear spreading with time {which occurs when perturbations are considerably weaker}.

\subsection{Three-body Divergence: Sun, Planet and Halley}\label{Sec:App2}


{We continue with the study of the effects of a single planet, but now we consider each planet of the Solar System in turn{, and not only Jupiter}.  Also we adopt the mass appropriate to each planet, without changing it artificially as in the previous section.
As in the three-body integrations of Sec.~\ref{sec:numerical},} 
we generate an ensemble of a thousand initial conditions, where we vary the initial orbital phase of that planet{, and for each initial phase we integrate two orbits for Halley, separated initially by $10^{-6}$ in one coordinate}.  {We once again use \texttt{Brutus} as the integrat{or}, but now} we integrate the system for 10\,kyr.
In {each different} 
integration, Halley will experience a different encounter history with the planet, which should produce different rates of divergence {within each orbit, in analogy with the results of the simplified model discussed in} 
Sec.~\ref{sec:sawtooth}.  
We show a subset of illustrative cases in Fig. \ref{fig:N3_sun_planet_halley_overview}{, which presents graphs of $\delta$ defined as in Eq.~(\ref{eq:norm})}.  

We first {consider} 
the results {for} 
Jupiter. The rates of divergence vary widely. There are solutions which stay almost constant within a time span of 10\,kyr (flat, yellow curve starting at $\log_{10} \delta = -6$). In the other extreme are solutions that grow exponentially and have ``saturated'' or completely diverged {(i.e. the separation of two orbits is limited by the size of Halley's orbit)} within a few thousand years (blue {and} 
green 
curves). In between, there are solutions with different kind of transitions in the divergence. After an initial flat phase of a certain duration, a transition to an exponential growth is possible (red and purple curves), but it is also possible for this exponential growth to {flatten off before complete divergence can be achieved} 
(cyan curve). 

The influence of Saturn on Halley's stability is less strong, but some solutions still grow exponentially for a few thousand years, after which they make a transition to 
{oscillatory behaviour.} 
The {divergence} 
of the perturbation never 
becomes {complete, in the sense introduced in the previous paragraph, at least in the time span of these integrations.} 
The slope in the exponential part of the blue curve is also shallower than the slope in {the exponential examples among} Jupiter's results. The remaining outer planets show a{n approximately linear} 
growth and thus have a negligible contribution to Halley's chaoticity.

The influence of the terrestrial planets varies. Mercury shows {approximately linear} 
behaviour irrespective of its encounter history with Halley.  
Venus on the other hand is able to produce fast growing solutions similar to  Jupiter. The most rapid{ly} growing solution saturates, i.e. the perturbation has become the size of the orbit, within 2\,kyr.  {For} Earth and Mars the majority of  solutions show a{n approximately linear}  
divergence superposed with {oscillatory behaviour}. 
Note, however, that they are able to generate a rapid growth in some situations. 

{For each planet, all} 
remaining solutions in the ensemble show a similar behaviour to {those} 
illustrated in Fig. \ref{fig:N3_sun_planet_halley_overview}. One statistic {which} we calculated is the fraction of rapidly growing solutions in the ensemble per planet. This gives an estimate of the likelihood that a planet is the dominant perturber of Halley. In Fig. \ref{fig:fraction_div} we plot the fraction of solutions that have reached saturation, {i.e. $\delta = 1$}, as a function of time. We confirm that Jupiter, Venus and Earth are the dominant perturbers of Halley, with relative fractions at 10\,kyr of $f_{div,Venus}/f_{div,Jupiter} = 0.68$ and $f_{div,Earth}/f_{div,Jupiter} = 0.28$. 

\subsection{Hopping Between Planets}\label{sec:hophop}

In this experiment we do not randomise the initial orbital phase, but we take the fiducial initial conditions such that we can measure the actual encounter histories of the planets with Halley. We consider the three-body systems including the Sun, a planet and Halley, to measure the independent rates of divergence. Based on the results of Sec.~\ref{Sec:App2}, we neglect Mercury, Uranus and Neptune. We compare these results with a simulation including all the relevant planets collectively. The results are given in Fig. \ref{fig:full_solar_system}{, which plots $\delta$ defined as in Eq.~(\ref{eq:norm})}. We averaged the data over bins of two orbital periods to reduce the short term oscillatory behaviour. 

We observe that only Venus (green curve ending at $\log_{10} \delta \sim 0.4$) and Jupiter (yellow curve crossing the curve of Venus at $\sim$ 3\,kyr) produce an exponential divergence. Initially the perturbation due to Venus dominates, but it is overtaken by Jupiter after about 3\,kyr due to a rapid sequence of close encounters { between Jupiter and Halley}. 

{More specifically, Halley will encounter Venus in 1019 years at a distance of 0.054\,AU, in 1317 years at 0.10\,AU, in 1514 years at 0.083\,AU and in 2296 years at 0.11\,AU. It is this sequence of close approaches that causes Venus to be the dominant perturber of Halley in the next few millennia. In the same time interval there are three close encounters between Halley and Jupiter, most notably {one} in 2607 years at 0.61\,AU. 
Following this
close encounter there is a higher density of close encounters between Halley and Jupiter, which causes the rapid exponential growth.} The solution including multiple planets (black curve) 
{exhibits} {a} 
transition {in which} 
it first follows the perturbations due to Venus and then hops onto the perturbations by Jupiter. 
Other 
effects are present since the black curve does not perfectly overly on top of the green and yellow curves. The superposition of independent growth rates is however a reasonable approximation in this example. 

{The validity of this approximate superposition is not necessarily to be expected.  In the integrations plotted in Fig.\ref{fig:full_solar_system}, the sequence of encounters in the full integration is different from that in any of the three-body integrations.   Since it appears from Fig.\ref{fig:map_sawtooth_overview} that the sequence of encounters is critical to transitions and the overall rate of divergence of neighbouring solutions, one might have expected that the contribution of Venus (for example) in the full integration could be quite different from that in the three-body integration in which Venus is the only perturber.  This expectation, however, does not appear to be borne out.}



{ In order to estimate the Lyapunov time scale of Halley's orbit, we perform a set of simulations similar to the one of the complete system in Fig. \ref{fig:full_solar_system} (black curve). We adopt the same method as \citet{2015MNRAS.447.3775M} and vary the direction of the initial perturbation in Halley's orbit to lie along the six different Cartesian axes (position and velocity), both in the plus and minus directions. The initial perturbation in position is set to $10^{-6}$ AU, and for the velocity to $4.4\times10^{-8}$ AU/yr\footnote{Since position and velocity have different units they require different initial offsets in order to produce a perturbation growth of similar magnitude.}. Together with the fiducial initial condition we obtain a set of 13 initial realisations. We integrate each system and subsequently measure the rate of divergence of each system compared to the fiducial solution. We regard the growth of a perturbation from t=0 until saturation of the perturbation when $\delta = 1$, and perform a simple linear regression with the initial offset fixed at $\delta_0 = 10^{-6}$ AU. The resulting Lyapunov time scales vary between 300 $\pm$ 1.6 and 335 $\pm$ 1.0 years with an average of 323 years. Our rough estimate of the minimal Lyapunov time scale of Halley's orbit is about 300 years, and thus considerably longer than its orbital period, as was the value found by \citet{2015MNRAS.447.3775M}. }
Finally, we also performed an experiment where we integrated backwards in time, to see when Venus became the dominant perturber. We find that {\sl both} Venus and Jupiter show a similar exponential divergence, reaching saturation between about -3 and -4\,kyr. The rate of divergence is asymmetric around the current time.  

\begin{figure}
\centering
\begin{tabular}{c}
\includegraphics[height=0.36\textwidth, width=0.45\textwidth]{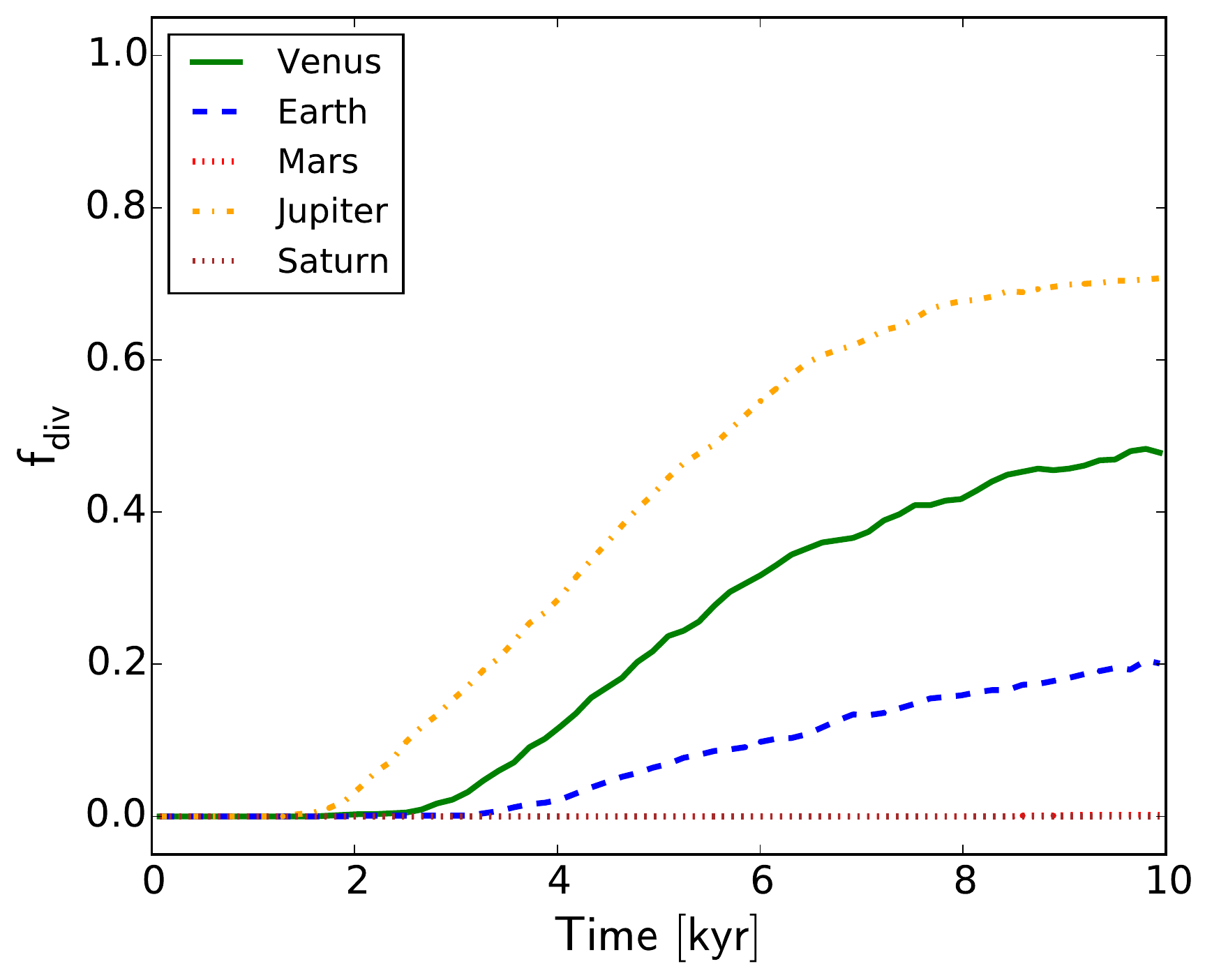} \\
\end{tabular}  
\caption{ The fraction of  solutions {which have} diverged up to saturation, $f_{div}$, as a function of time, for a subset of planets. We observe that Jupiter, Venus and Earth are the dominant perturbers of Halley. }
\label{fig:fraction_div}
\end{figure}

\begin{figure}
\centering
\begin{tabular}{c}
\includegraphics[height=0.36\textwidth, width=0.45\textwidth]{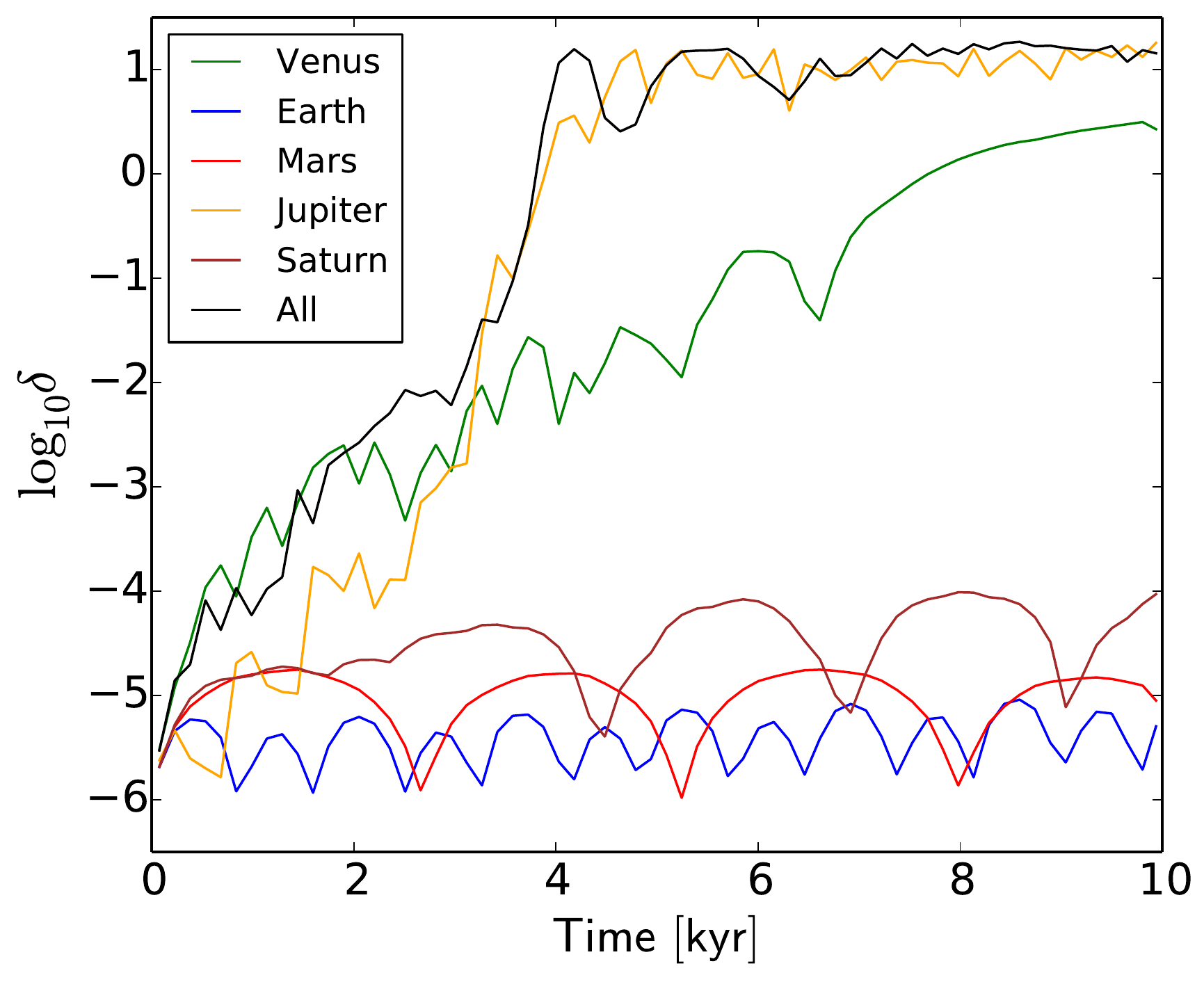} \\
\end{tabular}  
\caption{ Growth of perturbations in time for the different planets independently and with the planets collectively. Up to 3\,kyr, Venus is the dominant perturber of Halley's orbit. Then a transition occurs and Jupiter becomes the main perturber. The transition in the rate of divergence for the solution including all planets is explained {approximately} by the superposition of independent rates of divergence of {each of} the planets.  }
\label{fig:full_solar_system}
\end{figure}






\section{Discussion}\label{Sec:Discussion}
  

Previous studies have considered the value of 
the Lyapunov time scale for the growth of perturbations in Halley's orbit. \citet{2007IAUS..236...15S} gave an estimate of a lower bound of 34 years for the Lyapunov time scale and our estimate {(Sec.~\ref{sec:hophop})} is consistent with this.  Our estimate is, however, inconsistent with the results of \citet{2015MNRAS.447.3775M}, who found a value around 70 years.  This was based on an initial perturbation in the $y$-coordinate of Halley, but they also gave results for an initial perturbation in the $x$-coordinate (their Fig.~7) which would give a Lyapunov time scale only slightly longer.  We note, however, that their plot of the growth of the deviation between two orbits (their Fig.6) indicates growth in $\delta$ (their measure of the separation of two orbits) by about 5\,dex in 3.5\,kyr, implying a Lyapunov time scale of order 300 years, very similar to ours. One of the reasons for the discrepancy {with their published} 
value of the Lyapunov time scale could be the two different methods used to estimate the Lyapunov time scale. We measure the rate of exponential growth between two neighbouring trajectories directly until the moment the perturbation has saturated, while they use the iterative scheme from \citet{Benettin}. It is surprising to find that they give such different results, especially since both results are derived from a finite time integration of only a few thousand years.  

The Lyapunov time scale of Halley's Comet is determined principally by perturbations due to Venus and Jupiter (see Fig.\ref{fig:N3_sun_planet_halley_overview}). The influence of Earth, Mars and Saturn is smaller during the next few millennia, and that of Mercury, Uranus and Neptune is negligible. {Backward integrations showed that both Jupiter and Venus were dominant up till at least 3\,kyr in the past.} Generally, as expected, Jupiter takes the role of being the main perturber of Halley's orbit (see Fig. \ref{fig:fraction_div}). However, as {implied by the results of} 
Sec.~\ref{Sec:Encounter}, during a phase of relatively weak encounters with Jupiter, Venus can fuel the chaoticity of Halley's orbit instead.  

The comparable importance of Jupiter and Venus could not have been guessed from their relative masses {alone, and we showed in Sec.~\ref{sec:analytical-estimates} that}
the reason for this is {that} the contribution {of a given planet} also  depends on the distance of closest approach.  This is made apparent by the fact that the divergence caused by these two planets depends strongly on the initial phase (see Sec.~\ref{Sec:Encounter} and Fig.\ref{fig:N3_sun_planet_halley_overview}).  {The implication of this is that the growth rate, averaged over several encounters, depends on the sequence of encounters, and especially on the occurrence of close encounters.}   Indeed \citet{2015MNRAS.447.3775M} draw attention to a forthcoming relatively close encounter with Jupiter after about 3.4 kyr, and its influence is {noticeable also in our results} 
in Fig.\ref{fig:full_solar_system}.
  {In Sec.~\ref{sec:sawtooth}} we drew attention to the {possible} importance of a near-resonance in the motions of Halley and Jupiter, {and its importance for the sequence of weak and strong encounters and hence} the resulting growth of divergence between neighbouring orbits (Fig.\ref{fig:map_sawtooth_overview}). 
For different planets such configurations will occur at different {epochs}, 
{depending on the evolution of}
the orbits of the {planets, and in particular their periods.} 

Much of our focus in Sec.~\ref{Sec:Exp1} was on the parameter $\mu$, which measures the derivative of our kick function $K(\phi)$.  {This can be estimated in order of magnitude with the approach in Sec.~\ref{sec:analytical-estimates}, but} 
also{, with greater precision,} 
from the results of \citet{2014arXiv1410.3727R}, bearing in mind that their kick function $F(x)$ is the change (per perihelion passage) in twice the binding energy of Halley, as a function of $x = \phi/(2\pi)$.  For Venus the largest value of $\vert F'\vert$ occurs over a range of $x$ of order 0.1 in which $F$ decreases between values of about $\pm0.5\times10^{-4}$.  Thus we estimate $F' \simeq -10^{-3}$, and infer that $K' \simeq -10^{-5}$, though care has to be taken with the different units used in the two studies. This results in $\mu\simeq -6\times10^{-5}$. For the case that $\mu < 0$, {we recall from Sec.~\ref{mapeq}} 
that the eigenvalues of $A$ are always real, giving exponential growth.  When $-1\ll\mu<0$, the Lyapunov time scale can be estimated from 
  \begin{equation}
    T_{Lyapunov} \simeq \frac{1}{\sqrt{-\mu\omega_j}}.\label{eq:lt}
  \end{equation}
Using this equation we estimate that the corresponding Lyapunov time scale is of order 400 years.  This is of the correct order to account for the most rapid growth in Fig.\ref{fig:N3_sun_planet_halley_overview} (second panel), but it would only occur for phase values within a fairly narrow range.  For Jupiter, similar estimates give a Lyapunov time scale  an order of magnitude smaller, again over a similar, limited range of phases.    For Venus there is actually another larger range of phase with $K' <0$, but $\vert K'\vert$ is smaller than the estimate we have given, and the Lyapunov time scale correspondingly longer.  For both planets the magnitude of $K'$ is {generally} smaller than these upper limits, and so when $K'>0$ Halley remains in the regime of oscillatory ``growth'' (Sec.~\ref{Sec:Exp1}).  

When the phases are such that this occurs, it is interesting to note that these perturbations make Halley more stable {than if it had} 
no perturbations at all{, as already noted in Sec.~\ref{Sec:ensemble}.}
{ This mechanism applies
equally well to a swarm of bodies, which for example results from 
collisional fragmentation. If the velocities of the debris are much smaller
than their orbital velocities, one would expect that the swarm spreads
out linearly over time along the orbit similar to Kepler shearing. Instead, we find that in the 
regime of weak encounters this spreading has a sublinear or oscillatory behaviour and as a consequence, swarms can remain compact for a longer time. 
In Sec. \ref{Sec:ensemble} we varied the mass of Jupiter and we were able to model the weak 
encounter regime for 0.2 and 0.5 times the Jupiter mass. In Sec. \ref{Sec:App2} however, we observed
that even with Jupiter's actual mass, there are ``sublinear'' solutions for Halley's
stability, as long as close encounters are avoided (see Fig. \ref{fig:N3_sun_planet_halley_overview} bottom left panel). From Fig. \ref{fig:fraction_div} we estimate that close encounters with Jupiter are avoided in about 30 percent of the solutions on a time scale of at least 10 kyr. 
In the inner Solar System a close encounter with Jupiter will occur sooner or later, and
therefore signatures of sublinear spreading might only be found in relatively young
swarms. Other planets can also induce this type of weak encounters, depending on the mass of the planet and the orbital configurations as explained in Sec. \ref{Sec:Exp1}.  }

\section{Conclusions}\label{Sec:Conclusions}

{We confirm that} the orbit of Halley's Comet is chaotic \citep{1989A&A...221..146C, 2007IAUS..236...15S, 2014arXiv1410.3727R, 2015MNRAS.447.3775M}, {but find that the} 
Lyapunov time scale 
{is} of order 300 years (measured over approximately the next 4\,kyr). This value is significantly longer than the values determined in previous studies, which were of order the orbital period of Halley.
{ Our value of the Lyapunov time scale was obtained by measuring directly the phase space distance between a fiducial solution and a perturbed one, until the magnitude of the perturbation in position space reached the size of the system. We varied the direction of the initial perturbation to lie along the six Cartesian axes (position and velocity), both in the plus and minus directions, resulting in an ensemble of 13 solutions. We estimated the Lyapunov time scales using linear fits to the perturbation growths in log-linear space. In order to compare with previous studies, we ignored non-gravitational effects, such as sputtering during perihelion passages, which might influence the Lyapunov time scale of Halley's orbit. } 

The approximate exponential growth of perturbations in Halley's orbit has       {important} contributions {not only} from 
Jupiter{, as is already known, but also from} Venus{.  Indeed,} 
currently Venus is the dominant perturber{,} and Jupiter takes over {only after} 
about 3\,kyr from now. 
{This result does not rely only on numerical integrations, as we also use very simple order-of-magnitude estimates to show that the distance of closest approach to Venus can compensate for its low mass.}
This dependency on both the mass and the distance of the perturbing planet has the consequence that chaos strongly depends on the orbital configuration of the system. For example, minor bodies in the Solar System with larger perihelia than Halley, or a different eccentricity and inclination, will experience a different encounter sequence with the planets, and thus show different chaotic behaviour. The characterisation of the chaotic properties of a variety of orbits in the Solar System will increase our general understanding of dynamical chaos in planetary systems.   

The growth of perturbations in the orbit of Halley due to each {separate} planet 
has three modes: linear, 
oscillatory, and exponential, depending on the strength of the gravitational kicks the planet {imparts to} 
Halley.  An exponential growth is caused by a sequence of strong encounters of Halley with a planet, causing a short lived, but significant jump in the perturbation growth, while during the interval between such encounters the cometary motion {may be described as} 
{linear or oscillatory, i.e. as} ``quasi-regular'' \citep{1998CeMDA..70..181T}.  
On the other hand, {we also point out that} a sequence of weak encounters causes the growth of perturbations to behave {in an} oscillatory {fashion}, resulting in 
growth {which is slower than in the presence of still weaker perturbations}. This mechanism also applies to an ensemble of bodies orbiting in the Solar System, {so that the lifetime of such a system can be longer than expected from a linear growth. }

\section{Acknowledgements}

We thank Anna Lisa Varri and Adrian Hamers for fruitful discussions on chaos and Halley's orbit. We also thank Alessandro Morbidelli for insightful comments on Lyapunov time scales and the nature of chaotic orbits. 
The authors {are} also {grateful to} 
the referee for 
constructive feedback that {helped us} improve 
our manuscript. 
This work was supported by the Netherlands Research Council (NWO) and by the Netherlands
Research School for Astronomy (NOVA).  
Part of the numerical computations were carried out on the Little Green Machine at Leiden
University.


\bibliographystyle{mn2e} 
\bibliography{comet_halley.bib}      


\label{lastpage}

\end{document}